\documentclass[pra,reprint,twocolumn,superscriptaddress,showpacs,floatfix]{revtex4-1}
\usepackage{amsmath}
\usepackage{mathrsfs}
\usepackage{txfonts}
\usepackage{amssymb}
\usepackage{graphicx}
\usepackage{hyperref}
\usepackage{ulem}
\usepackage{xcolor} 
\usepackage{color}
\usepackage{overpic}
\usepackage{psfrag}
\usepackage{tabularx}
\usepackage{multirow}
\usepackage{array}
\usepackage{placeins}
\newcommand{\PreserveBackslash}[1]{\let\temp=\\#1\let\\=\temp}

\usepackage{bm}
\usepackage{dsfont}
\usepackage{mathtools}

\makeatletter

\begin{document}

\title{Symmetry group factorization and unitary equivalence among Temperley-Lieb integrable models}

\author{Huan-Qiang Zhou}
\affiliation{Centre for Modern Physics, Chongqing University, Chongqing 400044, The People's Republic of China}

\begin{abstract}
It is shown that there is a hidden connection between the two well-studied sequences of the Temperley-Lieb (TL) integrable models -- the $q$-state quantum Potts (QP) models at the self-dual points and  the staggered ${\rm SU}(n)$ spin-$s$ chains with $n=2s+1$ ($s \ge 1$), in addition to the uniform ${\rm SU}(2)$ spin-$1/2$ Heisenberg model. For each sequence, symmetry group factorization arises, in the sense that if $q$ is factorized into $q_1$ and $q_2$, then the $q$-state QP model is unitarily equivalent to a combined QP model with the symmetry group  ${\rm S}_{q_1} \times {\rm S}_{q_2}$ or  if  $n$ is factorized into $n_1$ and $n_2$, then  the staggered ${\rm SU}(n)$ spin-$s$ chain  with the symmetry group ${\rm SU}(n)$  is unitarily equivalent to a combined  staggered ${\rm SU}(n_1) \times {\rm SU}(n_2)$ spin chain with the symmetry group ${\rm SU}(n_1) \times {\rm SU}(n_2)$, valid for both ferromagnetic (FM) and antiferromagnetic (AF) cases. Moreover,  the FM (AF) staggered ${\rm SU}(n)$ spin-$s$ chain is unitarily equivalent to the AF (FM) $q$-state QP model with $q=n^2$, as long as the size of  the AF (FM) staggered ${\rm SU}(n)$ spin-$s$ chain is doubled. A combination of the two distinct types of unitary equivalences yields a family of models such that they are essentially identical, but appear in different guises. Some physical implications for unitary equivalence among different TL integrable models are clarified.

\end{abstract}
\maketitle

Quantum integrable models continue to be one of the main research themes in the investigation into quantum many-body physics. Actually, they play a significant role in diverse research areas, ranging from condensed matter to field theory in physics~\cite{faddeev} and from quantum groups to knot theory in mathematics~\cite{drinfeld,jones}. This partially rests on the fact that quantum integrable models share a common algebraic structure -- the Yang-Baxter equation~\cite{baxterbook},  which in turn is strongly tied with  some well-studied associative algebras over  the complex field $\mathbb{C}$, including the Temperley-Lieb (TL) algebra~\cite{tla,baxterbook,martin,levy,saleur,schutz,nichols}, the Hecke algebra~\cite{hecke} and the Birman-Murakami-Wenzl algebra~\cite{bwm1,bwm2} (for a comprehensive review, cf.~Ref.~\cite{isaev}). The underlying logic here is that  a given representation of these associative algebras yields a solution to the quantum Yang-Baxter equation. For instance, there is a systematic way to derive a solution to the quantum Yang-Baxter equation for a representation of the Temperley-Lieb algebra via the so-called Baxterization~\cite{jones1}.

A natural question arises as to whether or not there is any possible connection between different physical realizations of the same representation of such an associative algebra, with the TL algebra as the simplest example. Here  by different physical realizations we mean that they are realized in terms of different physical degrees of freedom, which act on different local Hilbert spaces on lattice sites in a one-dimensional lattice.
Indeed, as we shall show below, different physical realizations of the {\it same} representation yield different models with different explicit symmetry groups. Note that an explicit symmetry group is, by definition, generated by on-site symmetry operators (cf.~\cite{seifnashri} for a definition), which in turn are closely related to the notion of topological defects~\cite{fendley}.
Usually, one may anticipate that different symmetry groups lead to different degeneracies for a certain energy level, which is shared between two different physical realizations of the same representation of the TL algebra, an observation that has been implicitly assumed  to determine the energy gap between the ground state and the first excited state in Refs.~\cite{barber} (cf.~also Ref.~\cite{barber2}). However, imagine if there is a unitary equivalence among  various TL integrable models, then they are essentially identical, but appear in different guises, in the sense that an explicit symmetry group in one model becomes {\it hidden} in another model and vice versa, as a result of the fact that a unitary transformation involved is usually highly non-local, thus rendering on-site symmetry generators into non-on-site symmetry generators.

In this work, we attempt to address this intriguing question for  different physical realizations of the free-ends TL algebra. Here we focus on the two well-studied sequences of the TL integrable models -- the $q$-state quantum Potts (QP) models at the self-dual points ($q \ge 2$) and  the  staggered ${\rm SU}(n)$ spin-$s$ chains for $n \ge 3$, with $n= 2s+1$ ($s \ge 1$), in addition to the uniform ${\rm SU}(2)$ spin-$1/2$ Heisenberg model. Note that for the  staggered ${\rm SU}(n)$ spin-$s$ chain, when $n \ge 2$, the symmetry group is the uniform ${\rm SU}(2)$ group if the system size is odd and the  staggered ${\rm SU}(n)$ group if the system size is even~\cite{goldensu3,jesse}.  From now on, we always refer to the $q$-state QP model at the self-dual point ($q \ge 2$) as the $q$-state QP model and the staggered ${\rm SU}(n)$ spin-$s$ chain for $s \ge 1$ and the uniform spin-$1/2$ Heisenberg model as the ${\rm SU}(n)$ spin-$s$ chain ($s \ge 1/2$) for brevity. 

As it turns out, for each sequence,  a phenomenon coined as symmetry group factorization occurs, in the sense that if $q$ is factorized into $q_1 \ge 2$ and $q_2 \ge 2$, then the $q$-state QP model is unitarily equivalent to a combined QP model with the symmetry group  ${\rm S}_{q_1} \times {\rm S}_{q_2}$ or  if  $n$ is factorized into $n_1 \ge 2$ and $n_2 \ge 2$, then the ${\rm SU}(n)$ spin-$s$ chain  with the symmetry group ${\rm SU}(n)$  is unitarily equivalent to a combined spin chain with the symmetry group ${\rm SU}(n_1) \times {\rm SU}(n_2)$. Here we stress that this unitary equivalence arising from symmetry group factorization is valid for both ferromagnetic (FM) and antiferromagnetic (AF) cases.  In particular, we note that
the $q$-state QP model with $q=p^2$ is factorized into a combined QP model with the symmetry group  ${\rm S}_{p} \times {\rm S}_{p}$ ($p \ge 2$),  coined as the double  $p$-state QP model ($q \ge 2$), with the  double transverse-field Ising model (TFIM) as the simplest example when $p=2$.
Notably, a unitary equivalence even exists between the FM $q$-state QP models and the AF ${\rm SU}(n)$ spin-$s$ chains. Specifically,   the FM (AF) ${\rm SU}(n)$ spin-$s$ chain is unitarily equivalent to the AF (FM) $q$-state QP model with $q=n^2$, as long as the size of  the AF (FM) ${\rm SU}(n)$ spin-$s$ chain is doubled. Combining the two distinct types of unitary equivalences, we are led to a family of unitarily equivalent models among various TL integrable models.

{\it The free-ends TL algebra. -} Here we focus on the free-ends TL algebra. It is an  associative algebra over  the complex field $\mathbb{C}$, which is generated by the unit element $1$ and the  $N-1$ generators $U_j$ ($j=1,2, \cdots, L$)~\cite{tla,baxterbook,martin,levy}
\begin{align}
({\rm a})&~~\; U_j^2=\zeta \; U_j,\nonumber\\
({\rm b})&~~\; U_jU_{j\pm 1}U_j=U_j, 1 \le j \pm 1\le N, \nonumber\\
({\rm c})&~~\; [U_j, U_{k}]=0, |j-k|>1. \label{tla}
\end{align}
Here $\zeta$ is a real number and $N$ is a positive integer.
The generators  $U_j$ ($j=1,2,\ldots,N-1$) yield a Hamiltonian on a chain with free ends
\begin{equation}
H=\pm \sum_{j=1}^{N-1}U_j.    \label{hamsuN}
\end{equation}
Here the overall sign  represents either FM or AF coupling, depending on a specific model.
We stress that $\zeta$  is the {\it only} parameter to characterize a representation of the TL algebra, in the sense that different but unitarily equivalent realizations in some specific Hilbert spaces must share the same $\zeta$. 

There are a few well-known sequences of realizations of the TL algebra. One is  the $q$-state QP model at the self-dual point~\cite{martin,qpotts,pfeuty}.  The generators $U_j$ take the form
\begin{align}
	U_{2i} &= \frac {1} {\sqrt{q}} (1+ \sum_{\delta=1}^{q-1} R_{i}^{\delta}R_{i+1}^{q-\delta}), \nonumber \\
	U_{2i-1} &= \frac {1} {\sqrt{q}} (1+ \sum_{\delta=1}^{q-1} M_{i}^{\delta}).
	\label{potts}
\end{align}
Here $R$ is a diagonal $q \times q$ matrix, with the $\mu$-th entry being  $\exp [(\mu-1)\pi i/q]$ ($\mu=1,2,\ldots,q$), and $M$ is defined as
\begin{equation}
	M=\begin{bmatrix} 0 & I_{q-1} \\ 1 & 0
	\end{bmatrix},
\end{equation}
where $I_{q-1}$ is the $(q-1)\times (q-1)$ identity matrix.  Note that $\zeta= \sqrt q$ and $N=2L$, with $L$ being the system size.  
The model is AF or FM if the overall sign in the Hamiltonian (\ref{hamsuN}) is plus or minus.  It possesses the ${\rm S}_q$ symmetry group. 
In particular, when $q=2$, the model becomes the TFIM (at the self-dual point). The generators $U_j$ take the form
\begin{align}
	U_{2i} &= \frac {1} {\sqrt{2}} (1+ \sigma_i^z \sigma_{i+1}^z), \nonumber \\
	U_{2i-1} &= \frac {1} {\sqrt{2}} (1+ \sigma_i^x),
	\label{ising}
\end{align}
where $\sigma_i^x$ and $\sigma_i^z$ denote the Pauli matrices at the lattice site $i$. It possesses the ${\rm S}_2\approx {\rm Z}_2$ symmetry group generated by $\eta = \prod_i \sigma_i^x$. 

Another is the ${\rm SU}(n)$ spin-$s$ quantum chain for $s \ge 1/2$, with $n=2s+1$~\cite{barber}. The generators  $U_j$ take the form
\begin{equation}
	U_j=(-1)^{2s}\left[\frac{2^s}{(2s)!}\right]^2\prod_{S=1}^{2s}[X_j-\frac{1}{2}S(S+1)+s(s+1)].
	\label{staggered}
\end{equation}
Here $X_j=\textbf{S}_j \cdot \textbf{S}_{j+1}$ and $\textbf{S}_j=(S_{j}^x,S_{j}^y,S_{j}^z)$, where $S_{j}^x$, $S_{j}^y$ and $S_{j}^z$ represent the spin-$s$ operators at the $j$-th site. 
The above $U_j$ constitute a realization of the TL algebra (\ref{tla}), with $\zeta= n$ and $N=L$. Note that the model ( \ref{hamsuN}) with the above $U_j$ possesses staggered  ${\rm SU}(n)$ symmetry group~\cite{affleck}. Indeed,  the generators of this staggered  ${\rm SU}(n)$ symmetry group may be expressed in terms of  $S_{j}^x$, $S_{j}^y$ and $S_{j}^z$~\cite{goldensu3,jesse}. The model is FM or AF if the overall sign in 
the Hamiltonian (\ref{hamsuN}) is plus or minus.  Specifically, the simplest realizations are the ${\rm SU}(2)$ spin-$1/2$ Heisenberg model and the ${\rm SU}(3)$ spin-1 biquadratic model. More precisely, for $s=1/2$, we have 
\begin{equation}
	U_j=\frac{1}{2} - 2 \textbf{S}_j \cdot \textbf{S}_{j+1},
	\label{xxx}
\end{equation}
and  for $s=1$, we have 
\begin{equation}
	U_j= -1 + ( \textbf{S}_j \cdot \textbf{S}_{j+1})^2.
	\label{biquadratic}
\end{equation}
Here we remark that this sequence may be re-expressed in terms of the ${\rm SO}(n)$ generators as a biquadratic model~\cite{zhang}, which may be regarded as a special case of the  ${\rm SO}(n)$ bilinear-biquadratic model --  an extension of the familiar spin-$1$ ${\rm SO}(3)$ bilinear-biquadratic model~\cite{chubukov,fath}. Note that symmetry group factorization discussed above also works for this sequence (cf. Section~\ref{sonbb} in the Supplementary Material (SM)).

{\it A mathematical lemma. -} Suppose $U^1_j$ and $U^2_j$ are two realizations of the TL algebra with $\zeta_1$ and $\zeta_2$ in terms of physical degrees of freedom acting on their respective Hilbert spaces $	\mathscr{H}_1$ and $\mathscr{H}_2$, then  $U^1_j U^2_j$ also generate a  realization  of the TL algebra with $\zeta=\zeta_1 \zeta_2$, where the Hilbert space $\mathscr{H}$ is the tensor product $\mathscr{H}_1 \otimes \mathscr{H}_2$, namely $\mathscr{H}= \mathscr{H}_1 \otimes \mathscr{H}_2$. Here  $\mathscr{H}_1$ and $\mathscr{H}_2$ are defined as the tensor product spaces of their respective local Hilbert spaces 
$\mathscr{H}_{1j}$ and $\mathscr{H}_{2j}$ on lattice sites labeled by $j$, so $\mathscr{H}_j = \mathscr{H}_{1j} \otimes \mathscr{H}_{2j}$. The proof goes as follows.  Note that $U^1_j$ always commute with $U^2_k$ for any $j$ and $k$, given they act on two different Hilbert spaces $\mathscr{H}_1$ and $\mathscr{H}_2$. 
It follows that $U^1_j U^2_j$ satisfy the defining relations (a), (b) and (c) in Eq.\;(\ref{tla}). Hence $U^1_j U^2_j$ generate a  realization of the TL algebra, with $\zeta=\zeta_1 \zeta_2$.

This lemma thus offers a systematic means to construct a different realization from two known ones. In particular, it works for two copies of the {\it same} realization of the TL algebra, if one simply takes $U^2_j$ to be formally identical to $U^1_j$, as long as they are realized in terms of distinct physical degrees of freedom, which in turn act on two distinct Hilbert spaces.

{\it Symmetry group factorization. -} Now we turn to symmetry group factorization, which allows to establish a connection between various TL integrable models, as follows from the mathematical lemma above. For the $q$-state QP model,  if $q$, as a compound positive integer, is factorized into two integers  $q_1 \ge 2$ and $q_2 \ge 2$, namely $q= q_1 \times q_2$, then it follows that the $q$-state QP  model, with the symmetry group being the symmetric group ${\rm S}_q$, is factorized into a combined model consisting of the $q_1$-state QP  model and the $q_2$-state QP  model, with the symmetry group being ${\rm S}_{q_1} \times {\rm S}_{q_1}$, if one identifies   $U^1_j$ with that for the  $q_1$-state QP model and $U^2_j$ with that for the $q_2$-state QP model. 
Similarly,  for the ${\rm SU}(n)$ spin-$s$ chain,  if $n$, as a compound positive integer, is factorized into two integer  $n_1 \ge 2$ and $n_2 \ge 2$, namely $n= n_1 \times n_2$, then it follows that the ${\rm SU}(n)$ spin-$s$ chain  is factorized into a combined model consisting of the ${\rm SU}(n_1)$ spin-$s_1$ chain and the ${\rm SU}(n_2)$ spin-$s_2$ chain, with the symmetry group being ${\rm SU}(n_1) \times {\rm SU}(n_2)$ ($n_1=2s_1+1$ and $n_2=2s_2+1$), if one identifies  $U^1_j$ with that for the ${\rm SU}(n_1)$ spin-$s_1$ chain  and $U^2_j$ with that for the ${\rm SU}(n_2)$ spin-$s_2$ chain. Here the reduction of the symmetry group from ${\rm S}_q$ or ${\rm SU}(n)$ to  ${\rm S}_{q_1} \times {\rm S}_{q_1}$ or ${\rm SU}(n_1) \times {\rm SU}(n_2)$ reflects an intricate relation among different members in a unitarily equivalent family, thus justifying the introduction of explicit and hidden symmetry groups below.

One may keep repeating this symmetry group factorization for the $q$-state QP model or the ${\rm SU}(n)$ spin-$s$ chain  until one reaches a point that all factors of $q$ or $n$ are prime. According to the fundamental theorem in number theory, we have $q=q_1^{m_1} q_2^{m_2} \ldots q_\nu^{m_\nu}$ or $n=n_1^{m_1} n_2^{m_2} \ldots n_\nu^{m_\nu}$, where $\nu$ and $m_1,m_2,\ldots,m_\nu$ are positive integers,  and all of the factors $q_1, q_2, \ldots q_\nu$ or $n_1, n_2, \ldots n_\nu$ are prime. Many different realizations of the TL algebra are thus produced for the two sequences under investigation.

In particular, we are interested in the $q$-state QP model when $q$ is a perfect square $p^2$ ($p \ge 2$). 
Note that the $q$-state QP model is factorized into a combined QP model with the symmetry group  ${\rm S}_{p} \times {\rm S}_{p}$. The latter is coined as the double $p$-state QP model, with the double TFIM as the simplest example when $p=2$.
Consider two copies of the (critical) spin-$1/2$ TFIM in Eq.\;(\ref{ising}), which appear as a realization of the TL algebra:  $U^1_{j}(\sigma_i^x, \sigma_i^z)$ for $\sigma_i^x$ and $\sigma_i^z$, and $U^2_{j}(\tau_i^x, \tau_i^z)$ for $\tau_i^x$ and $\tau_i^z$, where $\tau_i^x$ and $\tau_i^z$ denote another set of the Pauli matrices at the lattice site $i$. As a result, we are led to a realization of the TL algebra:
\begin{align}
	U_{2i} &= \frac {1} {2} (1+ \sigma_i^z \sigma_{i+1}^z)  (1+ \tau_i^z \tau_{i+1}^z), \nonumber  \\
	U_{2i-1} &= \frac {1} {2} (1+ \sigma_i^x)(1+ \tau_i^x).
	\label{dising}
\end{align}
This realization shares the same $\zeta$ as the four-state QP model, with $\zeta = 2$. As already mentioned above, this model is coined as the double TFIM. It is readily seen that this model possesses the double Kramers-Wannier (KW) duality transformations, which inherit from the two copies of the TFIM (for the KW duality transformation, cf.~Refs.~\cite{levy,kadanoff}, adapted from the original form in the two-dimensional classical Ising model~\cite{kramers}). Indeed, this realization is a special integrable point of the quantum Ashkin-Teller model~\cite{ashkinteller1,ashkinteller2} (cf. Section~\ref{at} in the SM).

Similarly, if $n$ is a perfect square $n_p^2$ ($n_p \ge 2$), then the ${\rm SU}(n)$ spin-$s$ model is factorized into a combined spin-$s_p$ model ($n_p=2s_p+1$), with the symmetry group  ${\rm SU}(n_p) \times {\rm SU}(n_p)$. The latter is coined as the double ${\rm SU}(n_p)$ spin-$s_p$ model, with the double ${\rm SU}(2)$ spin-$1/2$ Heisenberg model as the simplest example when $n_p=2$. Consider two copies of the ${\rm SU}(n)$ spin-$s$ chain in Eq.\;(\ref{staggered}). One introduces $Y=\textbf{T}_j\cdot \textbf{T}_{j+1}$, where $\textbf{T}_j=(T_{j}^x,T_{j}^y,T_{j}^z)$, with $T_{j}^x$, $T_{j}^y$ and $T_{j}^z$ representing the second copy of the spin-$s$ operators at the $j$-th site. Then $U^1_j(X)U^2_j(Y)$ yields a realization of the TL algebra (\ref{tla}), with $\zeta= (2s_p+1)^2$.
Specifically, if $s_p=1/2$, then we have
\begin{equation}
	U_j=(\frac{1}{2} - 2 \textbf{S}_j \cdot \textbf{S}_{j+1}) (\frac{1}{2} - 2 \textbf{T}_j \cdot \textbf{T}_{j+1}),\label{hamst}
\end{equation}
and if $s_p=1$, then we have
\begin{equation}
	U_j=\left( 1 - ( \textbf{S}_j \cdot \textbf{S}_{j+1})^2 \right)
	\left(1 - ( \textbf{T}_j \cdot \textbf{T}_{j+1})^2 \right). \label{hamso4}
\end{equation}
Note that the double ${\rm SU}(2)$ spin-$1/2$ Heisenberg model (\ref{hamst}), with the symmetry group being ${\rm SU}(2) \times {\rm SU}(2)$, may be identified as a special point of the well-studied ${\rm SO}(4)$ spin-orbital model~\cite{so4}, and  the model (\ref{hamso4}) is  the double ${\rm SU}(3)$ spin-$1$  model, an extension of the spin-orbital model~\cite{so4} to the spin-1 case,  with the symmetry group being ${\rm SU}(3) \times {\rm SU}(3)$.

Our construction may be extended to any $m$ copies of  a given realization of the TL algebra. For the $q$-state QP model, we have $q=p^m$, including the cases that $q$ is a perfect cube when $m=3$ and $q$ is a perfect quadrangle when $m=4$. For the TFIM in (\ref{ising}), this extension yields the triple TFIM for $m=3$, with the symmetry group being ${\rm S}_2 \times {\rm S}_2  \times {\rm S}_2$,  and the quadruple TFIM for $m=4$, with the symmetry group being ${\rm S}_2 \times {\rm S}_2 \times{\rm S}_2  \times {\rm S}_2$.  For  the ${\rm SU}(2)$ spin-$1/2$ Heisenberg chain, we are led to the triple  spin-$1/2$ Heisenberg model for $m=3$ and the quadruple spin-$1/2$ Heisenberg chain for $m=4$, which may be re-interpreted as an extension of the spin-orbital model~\cite{so4}, with the symmetry groups being ${\rm SU}(2) \times {\rm SU}(2) \times {\rm SU}(2)$ and ${\rm SU}(2) \times {\rm SU}(2) \times {\rm SU}(2) \times {\rm SU}(2)$. 

Moreover, it is also possible to construct a realization of the TL algebra from two known realizations, one from the sequence of the $q$-state QP models and the other from the sequence of the ${\rm SU}(n)$ spin-$s$ chains. The simplest example describes an integrable version of the Ising-Heisenberg model, described by $U_j$ as follows 
\begin{align}
	U_{2i} &= \frac {1} {\sqrt{2}} (1+ \sigma_i^z \sigma_{i+1}^z) (\frac{1}{2} - 2 \textbf{S}_{2i} \cdot \textbf{S}_{2i+1}), \nonumber \\
	U_{2i-1} &= \frac {1} {\sqrt{2}} (1+ \sigma_i^x)  (\frac{1}{2} - 2 \textbf{S}_{2i-1} \cdot \textbf{S}_{2i}).
	\label{ising-heisenber}
\end{align}
Here we have chosen $U^1_j$ to be the generators (\ref{ising}) of the TL algebra for the TFIM and $U^2_j$ to be the generators (\ref{xxx}) of the TL algebra for the ${\rm SU}(2)$ spin-$1/2$ Heisenberg model.  The integrability of the Hamiltonian (\ref{ising-heisenber}) simply follows from the mathematical lemma above.
Note that the number of lattice sites is $L$ in the $\sigma$-sector, but the number of lattice sites is $2L$ in the $\textbf{S}$-sector. There are many other generalizations if one replaces the TFIM  and the ${\rm SU}(2)$ spin-$1/2$ Heisenberg model by the $q$-state QP model and the  the ${\rm SU}(n)$ spin-$s$ chain, respectively.

{\it Unitary equivalence. -} We resort to a mathematical proposition, stating that two given Hermitian Hamiltonians $H_1$ and $H_2$ are unitarily equivalent to each other, namely  $H_1 = V H_2 V^\dagger$, where $V$ is a unitary matrix, if and only if   ${\rm Tr}\; H_1^\kappa =  {\rm Tr} \;H_2^\kappa$ for all $\kappa \in \{ 0,1,2,\ldots,\infty \}$, where  $H_1$ and $H_2$ are defined in the Hilbert spaces $\mathscr{H}_1$ and $\mathscr{H}_2$, respectively. To the best of our knowledge, this proposition has not been formalized. Nevertheless, its implications are far-reaching, when one applies it to the two sequences of TL integrable models, as we shall show below. Here we sketch the proof as follows.  If $H_1 = V H_2 V^\dagger$ holds, then it is readily seen that  ${\rm Tr}\; H_1^\kappa =  {\rm Tr} \;H_2^\kappa$, due to the unitarity of $V$, namely $V V^\dagger = I_{\mathscr{H}_1}$ and $ V^\dagger V= I_{\mathscr{H}_2}$, and the cyclic property under the trace operation. Here $I_{\mathscr{H}_1}$ and $I_{\mathscr{H}_2}$ are the identity operators in $\mathscr{H}_1$ and $\mathscr{H}_2$. The converse is also true. This results from the observation that  ${\rm Tr}\; H_1^\kappa =  {\rm Tr} \;H_2^\kappa$ implies $\sum_{k_1} n^1_{k_1} (\epsilon^1_{k_1})^\kappa = \sum_{k_2} n^2_{k_2} (\epsilon^2_{k_2})^\kappa$, where $\{\epsilon^1_{k_1} \}$ and $\{\epsilon^2_{k_2} \}$ represent the entire spectra of $H_1$ and $H_2$, respectively, with  $n^1_{k_1}$  and $n^2_{k_2}$ being the degeneracies, thus representing the dimensions of the energy eigen-subspaces for $\{\epsilon^1_{k_1} \}$ and $\{\epsilon^2_{k_2} \}$. Here we have assumed that $\{\epsilon^1_{k_1} \}$ and $\{\epsilon^2_{k_2} \}$ are arranged in an increasing order, namely  $\epsilon^1_{k_1} > \epsilon^2_{l_1}$ if $k_1 > l_1$ and $\epsilon^2_{k_2} > \epsilon^2_{l_2}$ if $k_2 > l_2$.  As a consequence, $\{ k_1 \}$ and $\{ k_2 \}$ must be identified, denoted as $\{ k \}$, so that we are led to the identification: $\epsilon^1_k = \epsilon^2_k $ and $n^1_k=n^2_{k}$. Mathematically, this identification is attributed to infinitely many constraints imposed on the pair $\{\epsilon^1_{k_1}\}$ and $\{\epsilon^2_{k_2} \}$ and on the pair $\{n^1_{k_1}\}$ and $\{n^2_{k_2}\}$ for all possible $\kappa$.

Note that both $H_1$ and $H_2$ are Hermitian, so each of them may be diagonalized by performing a unitary transformation. Mathematically, this amounts to stating that both $H_1$ and $H_2$ admit a spectral decomposition $H_1= \sum_{k_1\alpha^1_{k_1}}\epsilon^1_{k_1} \vert k_1 \alpha^1_{k_1} \rangle \langle k_1 \alpha^1_{k_1} \vert$ and $H_2= \sum_{k_2 \alpha^2_{k_2}} \epsilon^2_{k_2} \vert \vert k_2 \alpha^2_{k_2} \rangle \langle k_2 \alpha^2_{k_2} \vert \vert$, where $\vert k_1 \alpha^1_{k_1} \rangle$ and  $\vert \vert k_2 \alpha^2_{k_2} \rangle$ represent (orthonormal) eigenstates of  $H_1$ and $H_2$ with the eigenvalues $\epsilon^1_k$ and $\epsilon^2_{k_2}$, and $\alpha^1_{k_1}=1,2,\ldots,n^1_{k_1}$ and $\alpha^2_{k_2}=1,2,\ldots,n^2_{k_2}$. If $H_1$ and $H_2$ are unitarily equivalent, then $\{ k_1 \}$ and $\{ k_2 \}$ are  identical, denoted as $\{ k \}$, so the unitary matrix $V$  connecting the two Hamiltonians $H_1$ and $H_2$ takes the form $V= \sum _{k\alpha_k} \vert k \alpha_k\rangle \langle k \alpha_k\vert \vert$ and $V^\dagger= \sum _{k\alpha_k} \vert \vert k \alpha_k\rangle \langle k \alpha_k\vert$ such that $V V^\dagger = \sum _{k\alpha_k} \vert k\alpha_k \rangle \langle k \alpha_k \vert$ and $V^\dagger V= \sum_{k\alpha_k} \vert \vert k \alpha_k\rangle \langle k \alpha_k\vert \vert$. Hence we have $V V^\dagger = I_{\mathscr{H}_1}$ and $ V^\dagger V= I_{\mathscr{H}_2}$, as follows from the resolution of the identity: $I_{\mathscr{H}_1}=\sum _{k\alpha_k} \vert k \alpha_k\rangle \langle k \alpha_k\vert$ and $I_{\mathscr{H}_2}=\sum _{k\alpha_k} \vert \vert k\alpha_k \rangle \langle k\alpha_k \vert \vert$ for $\mathscr{H}_1$ and $\mathscr{H}_2$.

A few remarks are in order. First, this mathematical proposition works for any two Hermitian Hamiltonians $H_1$ and $H_2$. In other words, they do not necessarily arise from any realization of the TL algebra (\ref{tla}). Here a small caveat is that, for a specific model, the Hamiltonian $H$ is only determined up to a positive multiplicative constant $\xi$ and an additive constant $\eta$, namely $H$ and $\xi H + \eta$ are physically equivalent. Hence we have to make sure that this freedom has been taken into account before the comparison between the pair
$\{\epsilon^1_{k_1}\}$ and $\{\epsilon^2_{k_2} \}$ is made. Second, for two Hamiltonians $H_1$ and $H_2$, if they constitute different realizations of the TL algebra, then it is necessary that they share the same $\zeta$ - the only parameter in the defining relations of the TL algebra (\ref{tla}). However, this alone is not sufficient to guarantee a unitary equivalence between $H_1$ and $H_2$.  Third,  if $H_1$ and $H_2$ are representations of the free-ends TL algebra with the same $\zeta$, then we are privileged to take advantage of this fact to drastically simplify the calculation of ${\rm Tr}\; H^\kappa$. Indeed, the free-ends TL algebra is finite-dimensional, in the sense that the number of words is finite, if one treats the generators $U_j$ as letters~\cite{levy,nichols}. This observation is very useful to establish a unitary equivalence between any two $H_1$ and $H_2$ arising from the TL algebra. Actually, when the above mathematical proposition is applied to TL integrable models, we are led to a mathematical corollary that
for two Hamiltonians $H_1$ and $H_2$ arising from the same sequence of the free-ends TL algebra via symmetry group factorization, they are unitarily equivalent if and only if they share the same partial trace of a reduced word, namely  ${\rm Tr}_{k,\chi}\;W^{(1)}_{k,\chi}={\rm Tr}_{k,\chi}\;W^{(2)}_{k,\chi}$, defined on a specific support consisting of $\chi$ adjacent lattice sites. For two Hamiltonians $H_q$ and $H_n$ arising from two different sequences of the free-ends TL algebra, they are unitarily equivalent if and only if  ${\rm Tr}_{k,\chi}\;W^q_{k,\chi}$ (up to a factor $q^{L-\chi}$ if $k+\chi-1$ is odd or a factor $q^{L-\chi-1}$ if $k+\chi-1$ is even) for the $q$-state QP model is identical to ${\rm Tr}_{k,\chi}\;W^n_{k,\chi}$ (up to a factor $n^{2L-\chi-1}$)  for the ${\rm SU}(n)$ spin-$s$ chain ($s \ge 1/2$), given that  the system size of the ${\rm SU}(n)$ spin-$s$ chain ($s \ge 1/2$) is doubled, compared to the size of the $q$-state QP model. In other words,  $q^{L-\chi}\;{\rm Tr}_{k,\chi}\;W^q_{k,\chi}$ 
if $k+\chi-1$ is odd or $q^{L-\chi-1}\;{\rm Tr}_{k,\chi}\;W^q_{k,\chi}$ if $k+\chi-1$ is even
for the $q$-state QP model must be identical to $n^{2L-\chi-1}{\rm Tr}_{k,\chi}\;W^n_{k,\chi}$ for the ${\rm SU}(n)$ spin-$s$ chain ($s \ge 1/2$).
(cf. Section~\ref{ue} in the SM). We are thus capable of establishing a family of unitarily equivalent models arising from the free-ends TL algebra.

{\it A family of unitarily equivalent models: explicit and hidden symmetry groups. -} A unitary transformation connecting two unitarily equivalent models among the same family is usually non-local, thus rendering an on-site symmetry group generator  into a non-on-site symmetry group generator (for a definition, cf. Ref.~\cite{seifnashri}). Hence it is necessary to introduce a notion -- a hidden symmetry group, which is usually generated by non-on-site symmetry generators, in contrast to an explicit symmetry group  generated by on-site symmetry generators.
In any circumstance, it is not an easy task to identify such a  (non-on-site) hidden symmetry group. However, the non-on-site nature alone is not sufficient to characterize a hidden symmetry group if one restricts to free-ends BCs, since various toroidal BCs also matter. In other words, unitary equivalence between two models is intricate so that the dichotomy between explicit and hidden symmetry groups, strictly speaking, needs to take different BCs into account.

According to the conventional wisdom, it is sufficient to focus on an explicit symmetry group once a specific model is given. However, this strategy could be misleading for two unitarily equivalent models. In fact, they may be naively thought of as related but different models, due to the fact that different explicit symmetry groups lead to different degeneracies for a certain energy level. However,
the existence of unitary equivalence among various TL integrable models invalidates the conventional wisdom, since it is possible to establish unitary equivalence between two different physical realizations of the TL algebra with the same $\zeta$ such that an explicit symmetry group for one model becomes a hidden symmetry group for another model and vice versa, if the unitary transformation connecting two members in a unitarily equivalent family is non-local.

If symmetry group factorization discussed above is taken into account, we are led to two distinct types of unitary equivalences for the two well-studied sequences of TL integrable models: one type is between two TL integrable models within each of the two sequences, and the other is between two TL integrable models from the two different sequences. Here we note that a few TL integrable models are introduced via symmetry group factorization (cf. Section~\ref{list} in the SM). As already mentioned above, the symmetry group from ${\rm S}_q$ or ${\rm SU}(n)$  for the $q$-state QP models or the ${\rm SU}(n)$ spin-$s$ model is reduced to  ${\rm S}_{q_1} \times {\rm S}_{q_1}$ or ${\rm SU}(n_1) \times {\rm SU}(n_2)$ for 
a combined model consisting of the $q_1$-state QP  model and the $q_2$-state QP  model or a combined model consisting of the ${\rm SU}(n_1)$ spin-$s_1$ chain and the ${\rm SU}(n_2)$ spin-$s_2$ chain. This reduction may be accounted for by resorting to explicit and hidden symmetry groups, thus
reflecting an intricate relation among different members in a unitarily equivalent family. Indeed, apparently different models from the same family are essentially {\it identical}, but they appear in different guises.

For instance,  the FM (AF) four-state QP model in Eq.~(\ref{potts}) with $q=4$, the AF (FM)  spin-$1/2$ Heisenberg model in Eq.~(\ref{xxx}) and the FM (AF) double TFIM in Eq.~(\ref{dising}) are in the same family, with the respective  (explicit) symmetry groups being ${\rm S}_4$, ${\rm SU}(2)$ and
${\rm S}_2 \times {\rm S}_2$. The FM (AF) nine-state QP model in Eq.~(\ref{potts}) with $q=9$, the AF (FM)  spin-$1$ ${\rm SU}(3)$ model in Eq.~(\ref{biquadratic}) and the FM (AF) double three-state QP model in Eq.\;(\ref{tpotts}) in the SM are in the same family, with the respective  (explicit) symmetry groups being ${\rm S}_9$, ${\rm SU}(3)$ and
${\rm S}_3 \times {\rm S}_3$. The FM (AF) eight-state QP model in Eq.~(\ref{potts}) with $q=8$, the FM (AF) triple TFIM in Eq.\;(\ref{tising}) in the SM and the Ising-Heisenberg model (\ref{ising-heisenber}) are in the same family, with the respective  (explicit) symmetry groups being ${\rm S}_8$, ${\rm S}_2 \times {\rm S}_2\times {\rm S}_2$ and  ${\rm S}_2 \times {\rm SU}(2)$. Moreover,  the FM (AF) sixteen-state QP model in Eq.~(\ref{potts}) with $q=16$, the FM (AF) quadruple TFIM in Eq.\;(\ref{tising}) in the SM,  and  the AF (FM) double spin-$1/2$ Heisenberg model in Eq.(\ref{hamst}) are in the same family. Note that the size of  the FM (AF) staggered ${\rm SU}(n)$ spin-$s$ model when $n=p$ must be doubled, compared to the sizes of  the AF (FM)  $q$-state QP model when $q$ is a perfect square $q=p^2$ and any other unitarily equivalent models that follow from symmetry group factorization, in order to ensure that the number of the TL generators $N-1$ must be the same (cf.~Section~\ref{ue} in the SM).

For all the models in the same family, they share all the symmetry groups, as long as one of the family members possesses such a specific symmetry group as explicit symmetry group. Note that this symmetry group is usually hidden for other members. As an example,  the FM (AF) four-state QP model possesses ${\rm S}_2 \times {\rm S}_2$ and ${\rm SU}(2)$ as hidden symmetries, the FM (AF) double TFIM possesses  ${\rm S}_4$ and ${\rm SU}(2)$ as hidden symmetries~\cite{ian}, and the AF (FM) spin-$1/2$ Heisenberg model possesses ${\rm S}_4$ and ${\rm S}_2 \times {\rm S}_2$ as hidden symmetries.  This explains why the ${\rm SU}(2)$ symmetry is observed in the FM four-state QP model~\cite{oshikawa} and why the ${\rm S}_4$ symmetry group appears in a special integrable point of the quantum Ashkin-Teller model~\cite{ashkinteller2,chepiga}, which is identical to the FM double TFIM.

{\it Physical implications. -} As is well-known, the FM spin-$1/2$ Heisenberg model undergoes spontaneous symmetry breaking (SSB) with one type-B Goldstone mode (GM), as follows from the counting rule~\cite{watanabe,hidaka}. Hence we are led to conclude that
one type-B GM also occurs in the AF four-state QP model and the AF double TFIM. One may anticipate that the same logarithmic scaling behavior of the entanglement entropy, as revealed for the FM spin-$1/2$ Heisenberg model~\cite{FMGM,popkov}, is valid for the AF four-state QP model and the AF double TFIM (at the self-dual points). Another conclusion one may draw from the unitary equivalence established here is that the ground state degeneracies for the AF $q$-state QP model with $q=p^2$ ($p > 2$) is identical to those for the FM ${\rm SU}(n)$ spin-$s$ chain with $n=p$, when its size is doubled. As shown in Refs.~\cite{goldensu3,jesse}, the ground state degeneracies are exponential with $L$.  More precisely, they are nothing but a subsequence of the Fibonacci-Lucas sequence (for a brief discussion, cf.~Section~\ref{ssb} in the SM).

In addition, the FM spin-$1/2$ Heisenberg model, the FM spin-$1$ ${\rm SU}(3)$ model and the FM ${\rm SU}(4)$ spin-orbital model (identified as the FM double ${\rm SU}(2)$ spin-$1/2$ Heisenberg model) are frustration-free (for the definition of a frustration-free Hamiltonian, cf.~Ref.\;\cite{tasakibook}). Hence the AF four-state QP model,  the AF nine-state QP model and the AF sixteen-state QP model are frustration-free. In fact, as shown in Ref.~\cite{zhougreen}, any Hamiltonian undergoing SSB with type-B GMs is frustration-free. Generically, the AF $q$-state QP model ($q =p^2$, with $p \ge 2$) undergoes SSB with $p-1$ type-B GMs, so it is frustration-free, given it is unitarily equivalent to the FM spin-$s$ ${\rm SU}(n)$ model with $n=p$. However, type-B GMs are {\it hidden}, since the broken symmetry group ${\rm SU}(n)$ is hidden.

{\it Summary and outlook. -} We have established various unitarily equivalent families among TL integrable models with free ends. In particular, symmetry group factorization  enables to reduce the construction of the KW duality transformation for the $q$-state QP model to a case when $q$ is a prime number, including $q=2$, $q=3$ and $q=5$ as the simplest examples.  Moreover, this makes it possible to investigate the KW duality transformations for many other TL integrable models, as long as they are in the same family as the $q$-state QP model. In particular, one may introduce the KW duality transformations for the ${\rm SU}(n)$ spin-$s$ chains, with the ${\rm SU}(2)$ spin-$1/2$ Heisenberg model and the ${\rm SU}(3)$ spin-$1$ biquadratic model as the two simplest examples.
Notably, this construction is necessary to derive their non-invertible KW duality symmetries~\cite{seiberg,shao1,shao,oshikawa}, which have been investigated for the TFIM under periodic and anti-periodic boundary conditions (BCs)~\cite{seiberg} and under various toroidal and non-toroidal BCs~\cite{zhou-ising}, the three-state QP models under periodic BCs~\cite{sierra} and under various various  toroidal and non-toroidal BCs~\cite{qianqianshi}  and the double TFIM under various  toroidal and non-toroidal BCs, which in turn is relevant to the four-state QP model and the ${\rm SU}(2)$ spin-$1/2$ Heisenberg model~\cite{zhou-doubleinsing}. To this end, a detailed investigation is also needed into unitary equivalences among various integrable models~\cite{qianshi2} arising from the periodic TL algebra~\cite{levy,nichols}.

In addition, Green parafermions have been realized (up to a projection operator) as emergent flat-band excitations in the FM spin-$1$ ${\rm SU}(3)$ model among many others~\cite{zhougreen}. As argued there, this construction is applicable to any models undergoing SSB with type-B GMs, as long as the ground state degeneracies are exponential with system size, but depend on what types of boundary conditions are adopted, as happens for the FM ${\rm SU}(n)$ spin-$s$ chains with $n=2s+1$ ($s \ge 1$). We thus anticipate that Green parafermions may be realized (up to a projection operator) in the AF $q$-state QP models  if $q > 4$.
In particular, if $q$ is a perfect square $p^2$ ($p > 2$), this follows from the unitary equivalence between the AF $q$-state QP model and the FM spin-$s$ ${\rm SU}(n)$ model with $n=p$, though Green parafermions are {\it hidden}. One may thus anticipate that there is an inherent connection between SSB with type-B GMs, Green parafermions and non-invertible KW duality symmetries in this family. We shall return to this intriguing topic in a forthcoming publication.

{\it Acknowledgment. -} I thank Murray T. Batchelor,  John O. Fj{\ae}restad, Ian P. McCulloch,  Jesse J. Osborne and Qian-Qian Shi for their collaboration  on a related project. I also thank Daniel Braak for carefully reading the manuscript and making comments to improve the presentation. 

{\it Note added. -} A mapping has been explicitly constructed to confirm that (i) the double TFIM and the ${\rm SU}(2)$ spin-$1/2$ Heisenberg model are unitarily equivalent, which may be extended to the quantum Ashkin-Teller model and the ${\rm U}(1)$ spin-$1/2$ XXZ chain; (ii) the triple TFIM and the integrable Ising-Heisenberg model are unitarily equivalent; (iii) the quadruple TFIM and the ${\rm SU}(4)$ spin-orbital model are unitarily equivalent~\cite{ian}.

\newpage
\section*{Supplementary Material}
\twocolumngrid
\setcounter{equation}{0}
\renewcommand{\theequation}{S\arabic{equation}}

\subsection{The ${\rm SO}(n)$ bilinear-biquadratic model}~\label{sonbb}

The ${\rm SO}(n)$ bilinear-biquadratic model is described by the Hamiltonian~\cite{zhang}, 
\begin{equation}
	\mathscr{H}(\theta)=\sum_{j}^L\left \{ \cos \theta \sum_{a<b}L_{j}^{ab}L_{j+1}^{ab}+\sin
	\theta \left( \sum_{a<b}L_{j}^{ab}L_{j+1}^{ab}\right) ^{2}\right \}.
	\label{hamibb}
\end{equation}
Here $L_{i}^{ab}$ denote the ${\rm SO}(n)$ generators at the lattice site $j$, and $\theta$ is used to parameterize the coupling constants.
When $\theta = \pi/2$ or $3\pi/2$, the Hamiltonian (\ref{hamibb}) becomes a biquadratic model, when it is expressed in terms of the ${\rm SO}(n)$ generators $L_{i}^{ab}$ at the lattice site $j$. Here $\theta = \pi/2$ and $\theta = 3\pi/2$ correspond to the FM and AF couplings. The symmetry group is uniform ${\rm SO}(n)$ group for odd $L$ and staggered ${\rm SU}(n)$ group for even $L$. Note that this biquadratic model is unitarily equivalent to the ${\rm SU}(n)$ spin-$s$ chain, and constitutes a realization of the TL algebra (\ref{tla}). 

For this sequence,  symmetry group factorization also works, in the sense that if $n$ is factorized into $n_1$ and $n_2$, then the ${\rm SO}(n)$ biquadratic model with the symmetry group ${\rm SO}(n)$ or ${\rm SU}(n)$  is unitarily equivalent to a combined model with the symmetry group being ${\rm SO}(n_1) \times {\rm SO}(n_2)$ or ${\rm SU}(n_1) \times {\rm SU}(n_2)$, if $L$ is odd or even. This is valid for both FM and AF cases. 

\subsection{The quantum Ashkin-Teller model}~\label{at}

The quantum Ashkin-Teller model~\cite{ashkinteller1,ashkinteller2} is described by the Hamiltonian

\begin{align}
	\mathscr{H}=& -J\sum_j (\sigma_i^z +\tau_i^z+ \Delta \sigma_i^z \tau_i^z) \nonumber  \\
	&-h\sum_j (\sigma_i^x \sigma_{i+1}^x + \tau_i^x \tau_{i+1}^x+ \Delta \sigma_i^x \sigma_{i+1}^x\tau_i^x \tau_{i+1}^x).
	\label{ashkinteller}
\end{align}
We remark that the Hamiltonian (\ref{ashkinteller}) is FM if  $h>0$ and AF if $h<0$, if $\Delta$ is fixed to be 1. For the FM  quantum Ashkin-Teller model,  a quantum phase transition occurs at a critical line $J=h$
between an ordered phase when $J$ is greater than $h$ and a disordered phase when $J$ is less than $h$~\cite{ashkinteller1} if one of the three coupling constants as an overall factor is fixed. It is in the Ashkin-Teller universality class with
exponents varying continuously with $\Delta$.

The Hamiltonian (\ref{ashkinteller}) with $J=h$ and $\Delta=1$ is identified with the  double TFIM in Eq.\;(\ref{dising}) (up to a (local) unitary transformation), which appears as a realization of the TL algebra, up to an additive constant. 
In addition to the double TFIM in Eq.\;(\ref{dising}), the trivial double TFIM consisting of two decoupled TFIMs corresponds to the Hamiltonian (\ref{ashkinteller}) when $\Delta=0$.

\subsection{Evaluation of ${\rm Tr}\; H^\kappa$ for  TL integrable models}~\label{ue}
 
Usually it is challenging to establish a unitary equivalence between two given Hermitian Hamiltonians $H_1$ and $H_2$. In contrast, it is relatively easy to show that two given Hermitian Hamiltonians $H_1$ and $H_2$ are not unitarily equivalent. According to the mathematical proposition stated in the main text, it is sufficient to demonstrate that two given Hermitian Hamiltonians $H_1$ and $H_2$ do not share the same eigenvalues. Usually, this can be checked analytically for a {\it specific} small system size $L$, in order to compare the pair
$\{\epsilon^1_{k_1}\}$ and $\{\epsilon^2_{k_2} \}$ and the pair $\{n^1_{k_1}\}$ and $\{n^2_{k_2}\}$.
In practice, one may perform exact diagonalization for a small system size $L$ to confirm that $H_1$ and $H_2$ are not unitarily equivalent, if the pair
$\{\epsilon^1_{k_1}\}$ and $\{\epsilon^2_{k_2} \}$ or the pair $\{n^1_{k_1}\}$ and $\{n^2_{k_2}\}$ are not identical.

For our purpose, we mainly concern two given Hamiltonians $H_1$ and $H_2$ that appear as different realizations of the TL algebra, with the same $\zeta$ - the only parameter in the defining relations (\ref{tla}) of the TL algebra. Here we emphasize that it is necessary for $H_1$ and $H_2$ to have the same $\zeta$, if they are unitary equivalent. However, the fact that they share the same $\zeta$ alone is not sufficient to guarantee their unitary equivalence. As already mentioned in the main text,  if $H_1$ and $H_2$ are two different realizations of the free-ends TL algebra with the same $\zeta$, then we are privileged to take advantage of this feature to drastically simplify the calculation of ${\rm Tr}\; H_1^\kappa$ and ${\rm Tr}\; H_2^\kappa$. This follows from the fact that the free-ends TL algebra is finite-dimensional, in the sense that the number of words is finite, if one treats the generators $U_j$ as letters to form words~\cite{levy,nichols}. Hence one only needs to classify different terms arising from an expansion of  $H_1^\kappa$
in terms of  generators $U^1_j$ and their counterparts for ${H_2}^\kappa$  to establish a unitary equivalence between $H_1$ and $H_2$.

A useful observation is that a pattern emerges for any possible contribution to  ${\rm Tr}\; H^\kappa$, when $H$ is expressed in terms of the $N-1$ generators $U_j$ (cf. Eq.\;(\ref{hamsuN})). A few remarks are in order. First, there are $\kappa!$ terms in total, each of which takes the form $U_{j_1}U_{j_2}\ldots U_{j_\kappa}$, where $j_\gamma \in \{ 1,2,\dots,N-1\}$, with $\gamma =1,2,\ldots,\kappa$. Second, it is always possible to use the defining relations in Eq.\;(\ref{hamsuN}) to simplify $U_{j_1}U_{j_2}\ldots U_{j_r}$ into a simpler form, which appears to be a product of various {\it reduced} words defined on non-overlapping supports, each of which consists of a few adjacent lattice sites. There are two types of reduced words, denoted as $W_{k,\chi}$ and
${\bar W}_{k,\chi}$: $W_{k,\chi}= U_k U_{k+1} \ldots U_{k+\chi-1}$ and ${\bar W}_{k,\chi}=U_{k+\chi-1}\ldots U_{k+1} U_k$. Here by a reduced word we mean it is impossible to be reduced into a simpler form by resorting to the defining relations (\ref{tla}) of the TL algebra (for a related but slightly different usage of this term, cf.~\cite{martin}). 
Hence the evaluation of  ${\rm Tr}\; H^\kappa$ is reduced to that of the partial trace of $W_{k,\chi}$, namely ${\rm Tr}_{k,\chi}\;W_{k,\chi}$. Here ${\rm Tr}_{k,\chi}$ denotes the partial trace that only acts on a support consisting of $\chi$ adjacent lattice sites labeled as $\{k,k+1,\dots,k+\chi-1\}$ if $k+\chi-1$ is odd or consisting of $\chi+1$ adjacent lattice sites labeled as $\{k,k+1,\dots,k+\chi\}$ if $k+\chi-1$ is even for the $q$-state QP model and on a support consisting of $\chi+1$ adjacent lattice sites labeled as $\{k,k+1,\dots,k+\chi\}$ for the ${\rm SU}(n)$ spin-$s$ chain ($s \ge 1/2$). Note that ${\rm Tr}(U_{j_1}U_{j_2}\ldots U_{j_\kappa})$ is a product of the partial traces of a few reduced words on non-overlapping supports, each of which takes the same form as either $W_{k,\chi}$ or  ${\bar W}_{k,\chi}$.  Third, it is convenient to take advantage of the fact that for all realizations of the TL algebra under investigation, $U_j$ is expressed in terms of traceless matrices, e.g., the Pauli matrices $\sigma_i^x$ and $\sigma_i^z$ or $R$ and $M$. This makes it much easier to evaluate ${\rm Tr}_{k,\chi}\;W_{k,\chi}$ and ${\rm Tr}_{k,\chi}\;{\bar W}_{k,\chi}$. However, one has to take an extra care of $S_{j}^x$, $S_{j}^y$ and $S_{j}^z$ for the ${\rm SU}(n)$ spin-$s$ chain ($s \ge 1$), since the terms consisting of $(S_j^z)^2$ also contribute to the partial traces.  Fourth,  it is necessary for any two unitarily equivalent Hamiltonians $H_1$ and $H_2$ to share the same $\zeta$, since the coefficients arising from this simplification in terms of  defining relations (\ref{tla}) of the TL algebra depend on $\zeta$. This condition ensures that a reduction of a generic word $U_{j_1}U_{j_2}\ldots U_{j_\kappa}$ to a product of reduced words produces the same coefficient for two different realizations.  Fifth,  for two Hamiltonians $H_1$ and $H_2$ arising from the same sequence, one may directly compare the partial traces ${\rm Tr}_{k,\chi}\;W^{(1)}_{k,\chi}$ and ${\rm Tr}_{k,\chi}\;{\bar W}^{(1)}_{k,\chi}$ with ${\rm Tr}_{k,\chi}\;W^{(2)}_{k,\chi}$ and ${\rm Tr}_{k,\chi}\;{\bar W}^{(2)}_{k,\chi}$. However, for two Hamiltonians $H_1$ and $H_2$ arising from two different sequences, it is necessary to take into account the different local Hilbert space dimensions and the system size doubling of the ${\rm SU}(n)$ spin-$s$ chain ($s \ge 1/2$). Sixth, ${\bar W}_{k,\chi}$ may be regarded as a spatial inversion of  $W_{k,\chi}$. When one evaluates the trace of $U_{j_1}U_{j_2}\ldots U_{j_\kappa}$, they always appear on different supports that are not overlapping with each other. Given their partial traces are identical, one may only focus on  $W_{k,\chi}$.

Putting everything together, we are led to a mathematical corollary that for two Hamiltonians $H_1$ and $H_2$ arising from the same sequence of the free-ends TL algebra via symmetry group factorization, they are unitarily equivalent if and only if they share the same partial trace of a reduced word, namely  ${\rm Tr}_{k,\chi}\;W^{(1)}_{k,\chi}={\rm Tr}_{k,\chi}\;W^{(2)}_{k,\chi}$, defined on a specific support consisting of $\chi$ adjacent lattice sites. For two Hamiltonians $H_q$ and $H_n$ arising from two different sequences of the free-ends TL algebra, they are unitarily equivalent if and only if  ${\rm Tr}_{k,\chi}\;W^q_{k,\chi}$ (up to a factor $q^{L-\chi}$ if $k+\chi-1$ is odd or a factor $q^{L-\chi-1}$ if $k+\chi-1$ is even) for the $q$-state QP model is identical to ${\rm Tr}_{k,\chi}\;W^n_{k,\chi}$ (up to a factor $n^{2L-\chi-1}$)  for the ${\rm SU}(n)$ spin-$s$ chain ($s \ge 1/2$), given that  the system size of the ${\rm SU}(n)$ spin-$s$ chain ($s \ge 1/2$) is doubled, compared to the size of the $q$-state QP model. In other words,  $q^{L-\chi}\;{\rm Tr}_{k,\chi}\;W^q_{k,\chi}$ 
if $k+\chi-1$ is odd or $q^{L-\chi-1}\;{\rm Tr}_{k,\chi}\;W^q_{k,\chi}$ if $k+\chi-1$ is even
for the $q$-state QP model must be identical to $n^{2L-\chi-1}{\rm Tr}_{k,\chi}\;W^n_{k,\chi}$ for the ${\rm SU}(n)$ spin-$s$ chain ($s \ge 1/2$).

As a consequence, the formidable task to establish a unitary equivalence between two TL integrable models is reduced to the evaluation of the partial trace of a reduced word $W_{k,\chi}$, which may be evaluated for the two sequences of TL integrable models under investigation. For the $q$-state QP model,
we have ${\rm Tr}_{k,\chi}\;W^q_{k,\chi}= q^{\chi/2}$ if $k+\chi-1$ is odd or ${\rm Tr}_{k,\chi}\;W^q_{k,\chi}= q^{\chi/2+1}$ if $k+\chi-1$ is even, and for a combined model consisting of the $q_1$-state QP  model and the $q_2$-state QP  model, we have  ${\rm Tr}_{k,\chi}\;W^{q_1q_2}_{k,\chi}= (q_1~q_2)^{\chi/2}$ if $k+\chi-1$ is odd or  ${\rm Tr}_{k,\chi}\;W^{q_1q_2}_{k,\chi}= (q_1~q_2)^{\chi/2+1}$ if $k+\chi-1$ is even. As a result, if $q=q_1 q_2$, then they are unitarily equivalent. For the  ${\rm SU}(n)$ spin-$s$ chain ($s \ge 1/2$), we have ${\rm Tr}_{k,\chi}\;W^n_{k,\chi}= n$. For a combined model consisting of the ${\rm SU}(n_1)$ spin-$s_1$ chain and the ${\rm SU}(n_2)$ spin-$s_2$ chain, we have  ${\rm Tr}_{k,\chi}\;W^{n_1n_2}_{k,\chi}= n_1n_2$.
As a result,  the ${\rm SU}(n)$ spin-$s$ chain is unitarily equivalent to this combined model, if $n=n_1n_2$. In addition,
according to the mathematical corollary above, we have  $q^{L-\chi}~{\rm Tr}_{k,\chi}~W^q_{k,\chi}=n^{2L-\chi-1}~{\rm Tr}_{k,\chi}~W^n_{k,\chi}$ if $k+\chi-1$ is odd or $q^{L-\chi-1}~{\rm Tr}_{k,\chi}~W^q_{k,\chi}=n^{2L-\chi-1}~{\rm Tr}_{k,\chi}~W^n_{k,\chi}$ if $k+\chi-1$ is even for two Hamiltonians $H_q$ and $H_n$ arising from two different sequences of the free-ends TL algebra. From this requirement one may deduce that  $q^{L-\chi/2}=n^{2L-\chi}$ for the $q$-state QP model with the size $L$ and the ${\rm SU}(n)$ spin-$s$ chain ($s \ge 1/2$) with the size $2L$. This implies that, if $n=p$ with $q=p^2$, then the AF or FM ${\rm SU}(n)$ spin-$s$ chain ($s \ge 1/2$) is unitarily equivalent to  the FM or AF $q$-state QP model, if the size of the former is doubled compared to the size of the latter. One may extend this evaluation to other realizations of the TL algebra arising from symmetry group factorization.
We are thus led to a few unitarily equivalent families among various TL integrable models, some of which have been enumerated in the main text. 

Here we remark that the necessity for doubling the size of  the FM (AF) staggered ${\rm SU}(n)$ spin-$s$ model, when it is compared to the size of  the AF (FM) $q$-state QP model, heavily relies on the fact that the number of the TL generators $N-1$ must be the same.
This amounts to requiring that the dimensions of the Hilbert spaces involved must be the same, as reflected in ${\rm Tr}\; H_1^\kappa =  {\rm Tr} \;H_2^\kappa$ when $\kappa =0$. Actually, ${\rm Tr}\; H^\kappa$ is nothing but the dimension of the Hilbert space $\mathscr{H}$ when $\kappa =0$.

We have briefly described an efficient way to evaluate ${\rm Tr}\;H^\kappa$ for realizations of the free-ends TL algebra. However, a modification is necessary to evaluate  ${\rm Tr}\;H^\kappa$ for realizations of the periodic TL algebra, since it is infinite-dimensional, as long as the number of the generators is greater than two~\cite{levy}. 
Once an efficient way to evaluate ${\rm Tr}\;H^\kappa$ is available for the periodic TL algebra, one may establish a family of unitarily equivalent models under toroidal boundary conditions~\cite{qianshi2}.

It is noteworthy to point out that a unitary transformation connecting two TL integrable models in the same family is usually highly non-local, thus rendering  (on-site) generators of an explicit symmetry group for one model to non-on-site generators of a hidden symmetry group for another. However, this is not always the case, as already encountered in the identification of the quantum Ashkin-Teller model (\ref{ashkinteller}) on the self-dual plane $J=h$ when $\Delta=1$ with the double TFIM in Eq.\;(\ref{dising}). Generically, this type of a (local) unitary transformation does not change the nature of a symmetry group. In other words, it maps an on-site symmetry operator into an on-site symmetry operator and vice versa. 

\subsection{Various TL integrable models via symmetry group factorization}~\label{list}
As discussed in the main text, even if one restricts to two well-studied sequences -- the $q$-state QP models and the ${\rm SU}(n)$ spin-$s$ chains ($s \ge 1/2$), we are capable of constructing  various TL integrable models via symmetry group factorization.  

The first model concerns $m$-copies of the  TFIMs. The Hamiltonian takes the form (\ref{hamsuN}), where the generators $U_j$ are as follows
\begin{align}
	U_{2i} &= \frac {1} { 2^{m/2}} \prod _{\alpha=1}^m (1+ \sigma_{\alpha,i}^z \sigma_{\alpha,i+1}^z), \nonumber  \\
	U_{2i-1} &= \frac {1} { 2^{m/2}}\prod _{\alpha=1}^m  (1+ \sigma_{\alpha,i}^x).
	\label{tising}
\end{align}
Here  $\sigma_{\alpha,i}^x$ and $\sigma_{\alpha,i}^z$ ($\alpha=1,2,\ldots,m$) denote $m$ sets of the Pauli matrices at the lattice site $i$.  Note that $\zeta = 2^{m/2}$ in Eq.\;(\ref{tla}). In particular, when $m=2$, Eq.\;(\ref{tising}) becomes Eq.\;(\ref{dising}) for the double TFIM, if one identifies $\sigma_{1,i}^x$ and $\sigma_{1,i}^z$  with  $\sigma_i^x$ and $\sigma_i^z$  and $\sigma_{2,i}^x$ and $\sigma_{2,i}^z$ with  $\tau_i^x$ and $\tau_i^z$. 
Moreover, the triple and quadruple TFIMs correspond to $m=3$ and $m=4$, respectively.

The second model consists of $m$-copies of the ${\rm SU}(2)$ spin-$1/2$ Heisenberg model models. The Hamiltonian takes the form  (\ref{hamsuN}), where the generators $U_j$ are as follows
\begin{equation}
	U_j=\prod _{\alpha=1}^m (\frac{1}{2} - 2 \textbf{S}_{\alpha,j} \cdot \textbf{S}_{\alpha,j+1}).\label{mhamst}
\end{equation}
Here  $\textbf{S}_{\alpha,j}=(S_{\alpha,j}^x,S_{\alpha,j}^y,S_{\alpha,j}^z)$, where $S_{\alpha,j}^x,S_{\alpha,j}^y$ and $S_{\alpha,j}^z$ represent $m$ sets of the spin-$1/2$ operators at the $j$-th site. Note that $\zeta = 2^{m}$ in Eq.\;(\ref{tla}). In particular, when $m=2$, Eq.\;(\ref{mhamst}) becomes Eq.\;(\ref{hamst}) for the double ${\rm SU}(2)$ spin-$1/2$ Heisenberg model, with the symmetry group ${\rm SU}(2) \times {\rm SU}(2)$. It has been identified as a special point of the well-studied ${\rm SO}(4)$ spin-orbital model~\cite{so4},
if one identifies $\textbf{S}_{1,j}=(S_{1,j}^x,S_{1,j}^y,S_{1,j}^z)$ with  $\textbf{S}_j=(S_{j}^x,S_{j}^y,S_{j}^z)$ and $\textbf{S}_{2,j}=(S_{2,j}^x,S_{2,j}^y,S_{2,j}^z)$ with  $\textbf{T}_j=(T_{j}^x,T_{j}^y,T_{j}^z)$, respectively. 
Note that the double, triple and quadruple spin-$1/2$ Heisenberg models correspond to $m=2$, $m=3$ and $m=4$, respectively.

The third model is decomposed into $m$-copies of the three-state QP models. The Hamiltonian takes the form  (\ref{hamsuN}), where the generators $U_j$  are as follows
\begin{align}
	U_{2i} &= \frac {1} {3^{m/2}} \prod _{\alpha=1}^m (1+ \sum_{\delta=1}^{q-1} R_{j}^{\delta}R_{i+1}^{q-\delta}), \nonumber \\
	U_{2i-1} &=\frac {1} {3^{m/2}} \prod _{\alpha=1}^m (1+ \sum_{\delta=1}^{q-1} M_{i}^{\delta}).
	\label{tpotts}
\end{align}
Here  $R_{i}^{\alpha}$ and $M_{i}^{\alpha}$ ($\alpha=1,2,\ldots,m$) denote $m$ sets of the Potts matrices at the lattice site $i$. Note that $\zeta = 3^{m/2}$ in Eq.\;(\ref{tla}). The double, triple and quadruple three-state QP models correspond to $m=2$, $m=3$ and $m=4$, respectively.\\

The list may be continued so that some types of realizations may be expected to be physically relevant to concrete problems in condensed matter.

\vspace{0.5cm}

\subsection{Ground state degeneracies and the Fibonacci-Lucas sequences for TL integrable models}~\label{ssb}

The FM spin-$1/2$ Heisenberg model is a paradigmatic example for SSB with type-B GMs~\cite{watanabe}, which is simultaneously the simplest realization of the TL algebra with a continuous symmetry group. Indeed,  many other TL integrable models exhibit  SSB with type-B GMs. For instance, the FM ${\rm SU}(3)$ spin-1 biquadratic model and the FM ${\rm SU}(4)$ spin-orbital model exhibit SSB with type-B GMs, as demonstrated in Refs.~\cite{goldensu3,jesse}. In particular,
they feature exponential ground state degeneracies with system size, thus yielding nonzero residual entropy.  Hence the unitary equivalence between the AF nine-state QP model, the AF double three-state QP model and the FM  spin-$1$ ${\rm SU}(3)$ model implies that their ground state degeneracies are identical. Similarly, the unitary equivalence between the AF sixteen-state QP model, the AF quadruple TFIM, and the FM double spin-$1/2$ Heisenberg model (or equivalently 
the FM  ${\rm SU}(4)$ spin-orbital model) implies that their ground state degeneracies are identical. 

As stressed in the main text, the size of  the FM staggered ${\rm SU}(n)$ spin-$s$ model must be doubled, compared to the size of  the AF $q$-state QP model when $q$ is a perfect square $p^2$ ($p \ge 2$), with $n=p$. Actually, this size doubling stems from the fact that there is a subtle difference between $N$  and $L$ for the two sequences investigated here. More precisely, $N=2L$ for
the $q$-state QP model and $N=L$ for the ${\rm SU}(n)$ spin-$s$ chain ($s \ge 1/2$). In other words, unitary equivalence between two TL integrable models from the two different sequences arises only when they have the same $N$, subject to the condition that they share the same $\zeta$ in the defining relations (\ref{tla}).

Generically, it follows that the ground state degeneracy ${\rm dim} \;\Omega^{\rm FE}_L(p)$ for the AF $q$-state QP model with $q=p^2$ is identical to that for the FM staggered ${\rm SU}(n)$ spin-$s$ model with $n=p$, when the size of the latter is doubled compared to the size of the former, if free-ends BCs are adopted. Note that free-ends BCs have been referred to as open BCs in Refs.~\cite{goldensu3,jesse} for the FM staggered ${\rm SU}(n)$ spin-$s$ model. Mathematically, we have ${\rm dim}\; \Omega^{\rm FE}_L(p)={\rm dim}\; \Omega^{\rm OBC}_{2L}(n)$, where ${\rm dim}\; \Omega^{\rm FE}_L(p)$ and ${\rm dim}\; \Omega^{\rm OBC}_{2L}(n)$ denote the dimensions of the ground state manifolds for 
the AF $q$-state QP model and the FM staggered ${\rm SU}(n)$ spin-$s$ model, with their sizes being $L$ and $2L$, respectively. Here we mention that this relation has been observed numerically in Ref.~\cite{qianshi2}.
In particular, if $p=2$, then we have  ${\rm dim}\; \Omega^{\rm FE}_L(p)=2L+1$ for the AF four-state QP model (cf.\;Eq.\;(\ref{ising}) with $q=4$) and the AF double TFIM (\ref{dising}); if $p >2$, then we have ${\rm dim}\; \Omega^{\rm FE}_L(p)=(R^{-4L-2}-R^{4L+2})/(R^{-2}-R^{2})$ for the AF $q$-state QP model,  where $R=(\!\sqrt{p+2}-\sqrt{p-2})/2$ as an extension of the (inverse) golden ratio, as follows from  the identification of ${\rm dim}\; \Omega^{\rm OBC}_{L}(n)$ with  the Fibonacci-Lucas sequence~\cite{goldensu3}. Note that ${\rm dim}\; \Omega^{\rm FE}_L(p)$ constitutes a subsequence of the Fibonacci-Lucas sequence.

\end{document}